\documentclass[twocolumn,showpacs,preprintnumbers,superscriptaddress]{revtex4}
\usepackage{amsmath}
\usepackage{amssymb}
\usepackage{amsmath}
\usepackage{epsfig}
\usepackage{graphicx}
\usepackage{inputenc}
\inputencoding{latin1}
\usepackage{ulem}

\ifx\hypersetup\sadfkjashdfkxja\else
\hypersetup{pdfsubject={}}
\hypersetup{pdfkeywords={}}
\hypersetup{plainpages=false}
\hypersetup{bookmarksnumbered=true}
\hypersetup{pdfstartview=FitH}
\hypersetup{pdfpagemode=UseNone}
\hypersetup{colorlinks=false}
\hypersetup{citebordercolor={.5 1 .5}}
\hypersetup{urlbordercolor={.5 1 1}}
\hypersetup{linkbordercolor={1 .7 .7}}
\fi

\newcommand{\ud}{\mathrm{d}}
\newcommand{\be}{\begin{equation}}
\newcommand{\ee}{\end{equation}}
\newcommand{\bea}{\begin{eqnarray}}
\newcommand{\eea}{\end{eqnarray}}

\newcommand{\Appendix}[1]%
    {%
     \section{#1}%
      }



\begin{document}

\title{Background field method in the large $N_f$ expansion of scalar QED}

\author{Zhi-Yuan Zheng}
\email{zhengzy@itp.ac.cn}
\affiliation{Key Laboratory of Theoretical Physics, Institute of Theoretical Physics, Chinese Academy of Sciences\\ Beijing 100190, People's
Republic of China}
\affiliation{School of Physical Sciences, University of Chinese Academy of Sciences\\No.19A Yuquan Road, Beijing 100049, China}
\author{Gai-Ge Deng}
\email{denggaige@gzhu.edu.cn}
\affiliation{Center for Astrophysics, Guangzhou University Guangzhou 510006, P.R.China}
\affiliation{ Guangzhou University Library, Guangzhou University Guangzhou 510006, P.R.China}
\begin{abstract}
Using the background field method, we, in the large $N_f$ approximation, calculate the beta function of scalar quantum electrodynamics at the first nontrivial order in $1/N_f$ by two different ways. In the first way, we get the result by summing all the graphs contributing directly. In the second way, we begin with the Borel transform of the related two point Green's function. The main results are that the beta function is fully determined by a simple function and can be expressed as an analytic expression with a finite radius of convergence, and the scheme-dependent renormalized Borel transform of the two point Green's function suffers from renormalons.
\end{abstract}
\maketitle
\section{Introduction}
\label{1.1}
In quantum field theory, the beta functions which determine the flows of the coupling constants are of fundamental importance. As is well-known, in the 1970s \cite{Gross:1973id,Politzer:1973fx}, it was the calculation of the beta function of a non-Abelian gauge theory (QCD) that led to the discovery of asymptotic freedom in this theory, which made theoretical physicists believe that this theory is the right theory for describing strong interactions. Since then, we have seen lots of efforts been put into calculating the beta functions of various theories, with the calculations of QED \cite{Kataev:2012rf,Baikov:2012zm,Baikov:2012rr} and QCD \cite{Herzog:2017ohr,Luthe:2017ttc,Baikov:2016tgj} having been calculated to five-loop order. In calculating the beta functions of gauge theories, the background field method which preserves the classical gauge invariance is an efficient method. In this method, we just need to calculate the related two point Green's functions for the background gauge fields \cite{Abbott:1980hw,Abbott:1981ke,Abbott:1983zw}.

Generally apart from the first few coefficients of the beta functions, we know little about them. Therefore it's meaningful to study the large order behaviour of quantum field theory under some approximation \cite{PalanquesMestre:1983zy,Gracey:1996tb,Holdom:2010qs,Alanne:2018ene}. An essential point, in the investigation of the large order behaviour of field theories, is whether the results obtained are convergent. The early investigations about this can be traced back to the works in Refs. \cite{Dyson:1952tj,Hurst:1952zk,Thirring:1953da}. In fact, our expressions obtained by perturbation methods, are generally at best asymptotic rather than  convergent series \cite{AD:2012kr}. The Borel transform, a mathematical technique, can be used to improve the convergence property of a series. To study the asymptotic behaviour of a series we can study its Borel transform which by definition has better convergence properties than the original series. After the acquirement of the Borel transform, if there are no singularities (renormalons), we can recover the original series \cite{Braaten:1998au,Beneke:1998ui,Beneke:1994sw,Beneke:1994qe}.

As is well-known, the beta functions provide us with useful information about the asymptotic behaviour of field theories, such as the asymptotic freedom in QCD (for a sufficiently large number of flavo\-u\-r we will lose this property). As regards the SM $U(1)$ gauge theory, its one-loop beta function suggests that it may suffer from a Landau pole which can be avoided if there is a nontrivial UV fixed point arising from the zero of the beta function. However, according to a lattice result given in Ref. \cite{Kogut:2005pm}, there is no nontrivial fixed point in a $U(1)$ gauge theory for $N_f=4$ \cite{Holdom:2010qs}. As has been shown in the literature \cite{PalanquesMestre:1983zy,Gracey:1996tb,Holdom:2010qs,Alanne:2018ene}, the large $N_f$ models can provide other possibilities; in Ref. \cite{PalanquesMestre:1983zy}, the beta function of spinor QED has been calculated at the leading order in $1/N_f$, and the result suggests that there might be nontrivial (UV and IR) fixed points. As regards the scalar QED, the positive three-loop beta function shown in Ref. \cite{Chetyrkin:1983qc} suggests that the running coupling increases monotonically towards the ultraviolet and thus will suffer from a Landau pole. Hence, analogous to spinor QED and QCD, it is meaningful to study the large $N_f$ behaviour of scalar QED---in this paper we shall calculate its beta function at the leading order in $1/N_f$ and discuss whether there are some possibilities to find some fixed points to avoid the Landau pole just mentioned before.

The remainder of this paper is organized as follows. In Sect. \ref{2}, we give a brief introduction to the background field method, and in Sect. \ref{3} derive the beta function in the background field method. In Sect. \ref{4}, we show the equivalence between two approaches of the background field method. In Sect. \ref{5}, we study the Borel transform of the two-point Green's function and derive an analytic expression for the beta function. Scheme dependence issues are discussed in Sect. \ref{6}. In Sect. \ref{7}, some numerical results about the beta function are given. Discussions and conclusions are presented in Sect. \ref{8}.
\section{A brief introduction to the background field method}
\label{2}
In this paper, we shall use the background field method to study scalar QED with the Lagrangian
\begin{equation}
\mathcal{L}(A,\phi)=-\frac{1}{4}(F_{\mu\nu}^0)^2+\mid D_{\mu}^0\phi^0\mid^2-m^2_0\mid \phi^0\mid^2,
\label{LoSQED}
\end{equation}
where $D_{\mu}^0=\partial_{\mu}+i e_0 A_{\mu}^0$ (through out this paper, the subscript or superscript $0$/$r$ in a quantity means that this quantity is a bare/renormalized quantity). This Lagrangian is invariant under transformations
\begin{equation}
\label{phitransfomation} \phi^0(x)\rightarrow \mathit {e}^{-i \alpha(x)}\phi_0,\qquad A^{\mu}_0(x)\rightarrow A^{\mu}_0(x)+\frac{1}{e_0}\partial _{\mu}\alpha(x).
\end{equation}
In ordinary quantum field formalism, the gauge invariance of scalar QED under transformations given in Eq. (\ref{phitransfomation}) is broken by the introduction of a gauge-variant gauge-fixing term.

Recalling that, to get the effective action $\Gamma$, we, in the background field formalism, can replace the quantum fields $\phi^0$ and $A^0$ in the conventional action with $\Phi^0=\phi^0+\phi_b^0$ and $\mathcal {A}^0=A^0+A_b^0$, $\phi_b^0$ and $A_b^0$ being the introduced background fields, and then use the formula
\begin{align}
\text{exp}\big\{i\Gamma[A_b^0\!,\!\phi_b^0]\big\}=&\int_{1PI}\mathcal {D}A^0\mathcal {D}\phi^0\mathcal {D}\phi^{*}_0\text{exp}\bigg\{i\int \ud x\nonumber\\&\Big[\mathcal{L}_{gf}(\mathcal {A}^0,A_b^0)\!+\!\mathcal {L}(\mathcal {A}^0,\Phi^0)\Big]\bigg\},
\label{sss}
\end{align}
where the subscript 1PI in Eq. (\ref{sss}) means that we include all diagrams, connected or not, each connected component being one-parti\-c\-le-irr\-e\-d\-u\-c\-ible \citep{Weinberg:1996kr}, and $\mathcal{L}_{gf}(\mathcal {A}^0,A_b^0)$ is the gauge-fixing term we choose in the background field method. Therefore, we can choose a gauge-fixing term which is invariant under the background gauge transformations given below in Eq. (\ref{gtsm0}) and Eq. (\ref{gtsm}) to preserve the gauge invariance of the effective action:
\begin{align}
A_{b,0}^{\mu}(x)&\rightarrow A_{b,0}^{\mu}(x)+\frac{1}{e_0}\partial^{\mu}\alpha(x),\quad A^{\mu}_0(x)\rightarrow A^{\mu}_0(x),\label{gtsm0}\\
\phi_b^0(x)&\rightarrow e^{-i\alpha(x)}\phi_b^0(x),\qquad\phi^0(x)\rightarrow e^{-i\alpha(x)}\phi^0(x).\label{gtsm}
\end{align}
The gauge-fixing term we choose is
\begin{equation}
\mathcal{L}_{gf}(\mathcal {A}^0,A_b^0)=-\frac{(\partial_{\mu}A^{\mu}_0)^2}{2\alpha_0}.
\label{gfix}
\end{equation}

The gauge invariance of the effective action guarantees that its divergence part, which is just a functional of the background fields, takes the form \citep{Weinberg:1996kr}
\begin{equation}
\Gamma[A_b^0,\phi_b^0]=\int \Big\{L_{\phi}\big|D^{\mu}_{b,0}\phi_b^0\big|^2-\frac{L_A}{4}(F^{\mu\nu}_{b,0})^2-L_mm^2_0\big| \phi_b^0\big|^2\Big\} \ud x\nonumber.
\end{equation}
Adding this divergent part to the classical part $\int \ud x\mathcal {L}(A_b^0,\phi_b^0)$, defining the renormalized quantities by
\begin{align}
\label{dyr1}A^{r}_b&=\sqrt {1+L_A} A^{0}_b=Z_A^{-\frac{1}{2}}A^{0}_b\\
\label{dyr2}\phi^r_b&=\sqrt {1+L_{\phi}}\phi_b=Z_{\phi}^{-\frac{1}{2}}\phi^0_b,\\
\label{dyr3}m_r^2&=\frac{(1+L_m)m^2_0}{1+L_{\phi}}.
\end{align}
we get the renormalized effective action
\begin{equation}
\Gamma^r[A^r_b\!,\!\phi^r_b]=\int dx \bigg[\big| \partial_{\mu}\phi^r_b\!+\!ie_0A^{\mu}_b\phi^r_b\big|^2\!-\!\frac{(F^{\mu\nu}_{b,r})^2}{4}\!-\!m_r^2\mid \phi^r_b\mid^2\bigg].\nonumber
\label{ref}
\end{equation}
All these quantities, except $e_0$ and $A_b$, appearing in this expression are now renormalized quantities. Therefore $e_0 A_b$ must be finite, i.e. if we define $e_0=Z_e e$, we can set
\begin{equation}
Z_e\sqrt {Z_A}=1.
\label{sinf}
\end{equation}

The treatment given above follows the presentation given in Ref. \citep{Weinberg:1996kr}. We can also derive out Eq. (\ref{sinf}) following the treatment given in Refs. \cite{Abbott:1980hw,Abbott:1981ke,Abbott:1983zw}: Since we have chosen a gau\-g\-e-fixing term being invariant under the background gauge transformations (in scalar QED there is no need to introduce the ghost field), the explicit gauge invariance is retained in the background field method. Therefore the infinities appearing in the gauge-invariant effective action $\Gamma[A_b\!,\!\phi_b]$ must be proportional to $|\phi^r_b|^2$, $|D^{\mu}_{b,r}\phi^r_b|^2$ and $[F^{\mu\nu}_{b,r}]^2$. Now, according to Eq. (\ref{dyr1}-\ref{dyr2}), $|D^{\mu}_{b,0}\phi^0_b|^2$, a tree order term of the effective action, can be written as
\begin{equation}
|D^{\mu}_{b,0}\phi^0_b|^2\!=\!|\partial_{\mu}\phi_b^0\!+\!i e_0 A_{b,0}^{\mu}\phi_b^0|^2\!=\!Z_{\phi}|\partial_{\mu}\phi_b^r\!+\!i e Z_e\sqrt{Z_A} A_{b,r}^{\mu}\phi_b^r|^2,\nonumber
\end{equation}
which will be proportional to $|D^{\mu}_{b,r}\phi^r_b|^2=|\partial_{\mu}\phi^r_b+i e A_{b,r}^{\mu}\phi^r_b|^2$ only if Eq. (\ref{sinf}) holds.

\section{Beta function in the background field method}
\label{3}
Through out this paper we will use the dimensional regularization (DR) procedure in $4-2\epsilon$ dimensions and choose the minimal subtraction (like) scheme---issues about the scheme dependence will be discussed in Sect. \ref{6}.

In DR the bare and renormalized couplings are related by $e_0=\mu^{\epsilon}eZ_e$, where $\mu$ is the renormalization scale. Using $Z_e\sqrt {Z_A}=1$ and the independence of the bare coupling $e_0$ on $\mu$, we have
\begin{equation}
\beta_{\epsilon}(e)\Big(2-e\frac{\partial}{\partial e}\Big)Z_A=-2\epsilon e Z_A,
\label{beta}
\end{equation}
where  $\beta_{\epsilon}(e)=-\epsilon e+\beta (e)$, and $Z_A$, in the minimal subtraction scheme, is written as a series of poles in $\epsilon$:
\begin{equation}
Z_A=1+\sum_{i=1}^{\infty}\frac{Z_A^{(i)}}{\epsilon^i}.
\label{ZA}
\end{equation}
Substitution of these two expressions into Eq. (\ref{beta}) gives
\begin{equation}
\beta(e)\Big(2-e\frac{\partial}{\partial e}\Big)Z_A=-\epsilon e^2\frac{\partial Z_A}{\partial e}.
\end{equation}
The various $Z_A^{(i)}$, according to this equation, are related by
\begin{equation}
\beta (e)\Big(2-e\frac{\partial}{\partial e}\Big)Z_A^{(i)}=-e^2\frac{\partial}{\partial e} Z_A^{(i+1)},
\label{rlation}
\end{equation}
and the beta function can be obtained by setting $i=0$:
\begin{equation}
\beta(e)=-\frac{1}{2}e^2\frac{\partial}{\partial e}Z_A^{(1)}.
\label{zbeta}
\end{equation}
In our large $N_f$ approximation, Eq. (\ref{rlation}) can be simplified to
\begin{equation}
\beta_1\big(j-1\big)Z_A^{(i,j)}=e(j+1)Z_A^{(i+1,j+1)},
\label{bgx}
\end{equation}
where $\beta_1$ is the one-loop beta function and $Z_A^{(i,j)}$ is the first order (order $1/N_f$) contribution of the $j$-loop diagrams to $Z_A$. This equation will act as a strong check on our calculations.
\section{Application of the background field method}
\label{4}
In this work, the renormalization constants for the quantum fields and the gauge parameter are defined by
\begin{align}
\label{d1}
A^{0}(x)\!=\!\sqrt {Z_3}A^r(x)\!,\!\quad\phi^0(x)\!=\!\sqrt{Z_{2}}\phi^r(x)\!,\!\quad\alpha_0\!=\!Z_{\alpha}\alpha.
\end{align}
As in the case of spinor QED, in scalar QED we can prove $Z_3=Z_{\alpha}$ by using the Ward identity (a brief proof is given in \ref{APD}). Because of $Z_3=Z_{\alpha}$, we have two distinct ways to carry out our calculations, which here are called ``direct approach'' and ``indirect approach'' respectively  \citep{Capper:1982tf}.

In the ``direct approach'', using  $Z_3=Z_{\alpha}$, we can cancel all the renormalization factors in the gauge-fixing term shown in Eq. (\ref{gfix}) and get
\begin{equation}
\mathcal{L}_{gf}=-\frac{\big(\partial^{\mu}A^r_{\mu}\big)^2}{2\alpha}.
\label{Dgf}
\end{equation}
In the ``indirect approach'', the gauge-fixing term is split into
\begin{equation}
\mathcal{L}_{gf}=-\frac{\big(\partial^{\mu}A_{\mu}^0\big)^2}{2\alpha}-\Big(\frac{1}{Z_{\alpha}}-1\Big)\frac{\big(\partial^{\mu}A_{\mu}^0\big)^2}{2\alpha}.
\label{IDgf}
\end{equation}

In the background field method, since each scalar propagator contributes a factor of $Z_{2}^{-1}$ and each corresponding vertex contributes a factor of $Z_{2}$, we can avoid the renormalization procedure for the complex scalar field $\phi$. In the ``direct approach'', since we have eliminated all the renormalization factors in the new gauge-fixing term shown in Eq. (\ref{Dgf}), we can't avoid the renormalization procedure for the photon field. This new gauge-fixing term in the ``direct approach'' will be combined with $-(F_{\mu\nu}^r)^2/4$ to generate the conventional photon propagator
\begin{equation}
D_F^{\mu\nu}(k)=\frac{-i\big(g^{\mu\nu}k^2-(1-\alpha)k^{\mu\nu}\big)}{(k^2)^2},
\label{cbz}
\end{equation}
and the corresponding counterterms is generated from $(1-Z_3)(F_{\mu\nu}^r)^2/4$.

In  the ``indirect approach'' where we can avoid the renormalization procedure for the photon field, the first term in the new gauge-fixing term shown in Eq. (\ref{IDgf}) can be used together with $(F_{\mu\nu}^0)^2/4$ to generate the ``bare'' photon propagator like that shown in Eq. (\ref{cbz}), while the second term can be used to generate a new photon-photon vertex $AA$ with the Feynman rule
\begin{equation}
\tilde {D}^{\mu\nu}(k)=-i\Big(\frac{1}{Z_{\alpha}}-1\Big)\frac{k^{\mu}k^{\nu}}{\alpha}.
\label{Nvertex}
\end{equation}
In deriving this Feynman rule and the ``bare'' photon propagator, we use the bare photon field instead of the usual renormalized photon field.
\subsection{Direct Approach}
\label{3.1}
\begin{figure}[t]
\begin{center}
\includegraphics[scale=0.45]{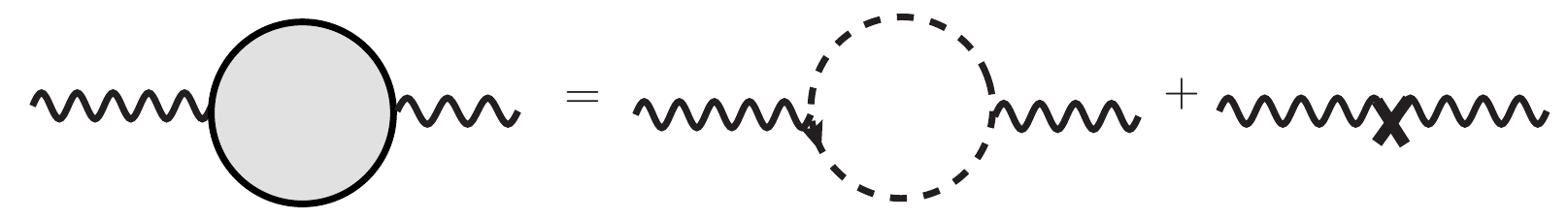}
\caption{One loop self-energy graph (renormalized, unrenormalized, counterterms)}
\label{xyz}
\end{center}
\end{figure}

Now, let's begin with the evaluation of $Z_3$ at one-loop level. Since we are concerned with the renormalization constants, we shall set the mass $m$ to zero. The diagram we should calculate is the unrenormalized diagram shown in Fig. \ref{xyz} where the scalar loop represents the total contribution of the $N_f$ charged spinless fields, with the result being
\begin{equation}
-i(g^{\mu\nu} k^2-k^{\mu}k^{\nu})\Big(\frac{4\pi \mu^2}{-k^2}\Big)^{\epsilon}\frac{e^2 N_f}{48 \pi^2}\frac{F(\epsilon)}{\epsilon},
\label{oneloop}
\end{equation}
where $k$ is the external momentum and
\begin{equation}
F(\epsilon)=\frac{3 \Gamma (1-\epsilon )^2 \Gamma (\epsilon +1)}{(3-2 \epsilon ) \Gamma (2-2 \epsilon )}.
\label{F}
\end{equation}
Therefore, the renormalization constant $Z_3$ at one loop level is given by
\begin{equation}
Z_3^1=-\frac{g}{\epsilon},
\end{equation}
with
\begin{equation}
g=\frac{e^2N_f}{48 \pi^2}.
\end{equation}
The zero order ($1/N_f^0$) diagram calculated above and its counterterms play a fundamental role in the $1/N_f$ expansion; they can be inserted into any diagram without changing the order in  $1/N_f$.

\begin{figure}[t]
\begin{center}
\includegraphics[scale=0.35]{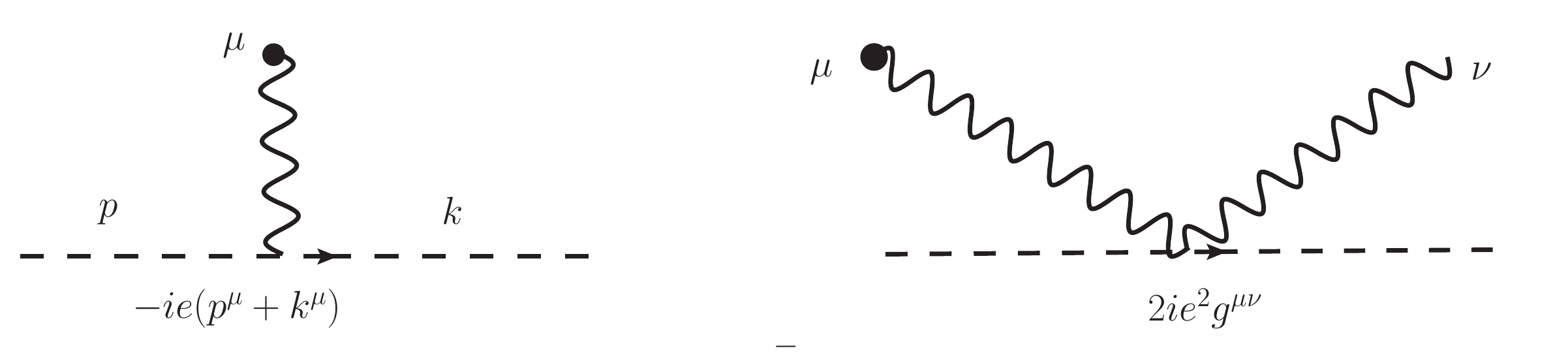}
\caption{Here the black bubbles in the end of the lines indicate that these lines are background field external legs.}
\label{ddian}
\end{center}
\end{figure}

Now, we turn to the evaluation of $Z_A$. The vertices involving background fields and their Feynman rules are given in Fig. \ref{ddian}. According to these Feynman rules, we can show that $Z_3^1=Z_A^1$. This identity and $Z_e\sqrt {Z_A}=1$ indicate that $Z_e\sqrt {Z_3}=1$ holds at one-loop level. Therefore, in our large $N_f$ approximation, we need not worry about the vertex corrections.
\begin{figure}[htb]
\centering
\includegraphics[scale=0.26]{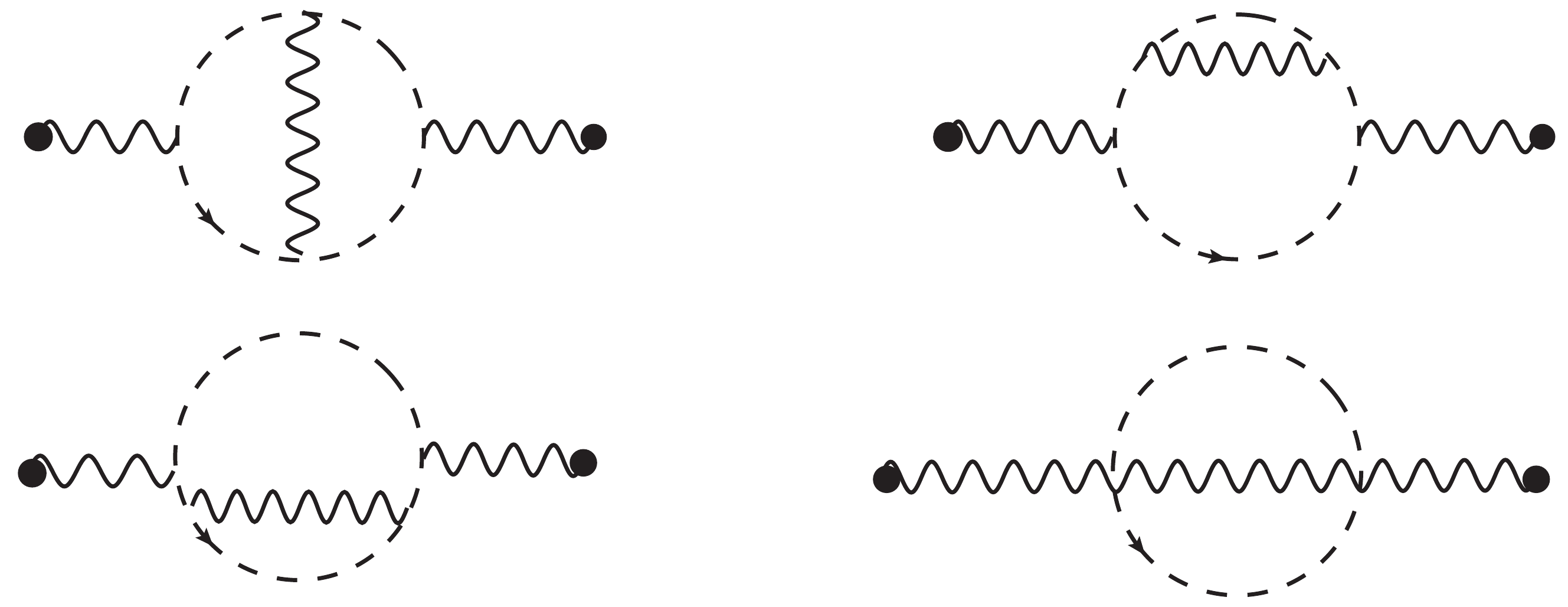}
\includegraphics[scale=0.26]{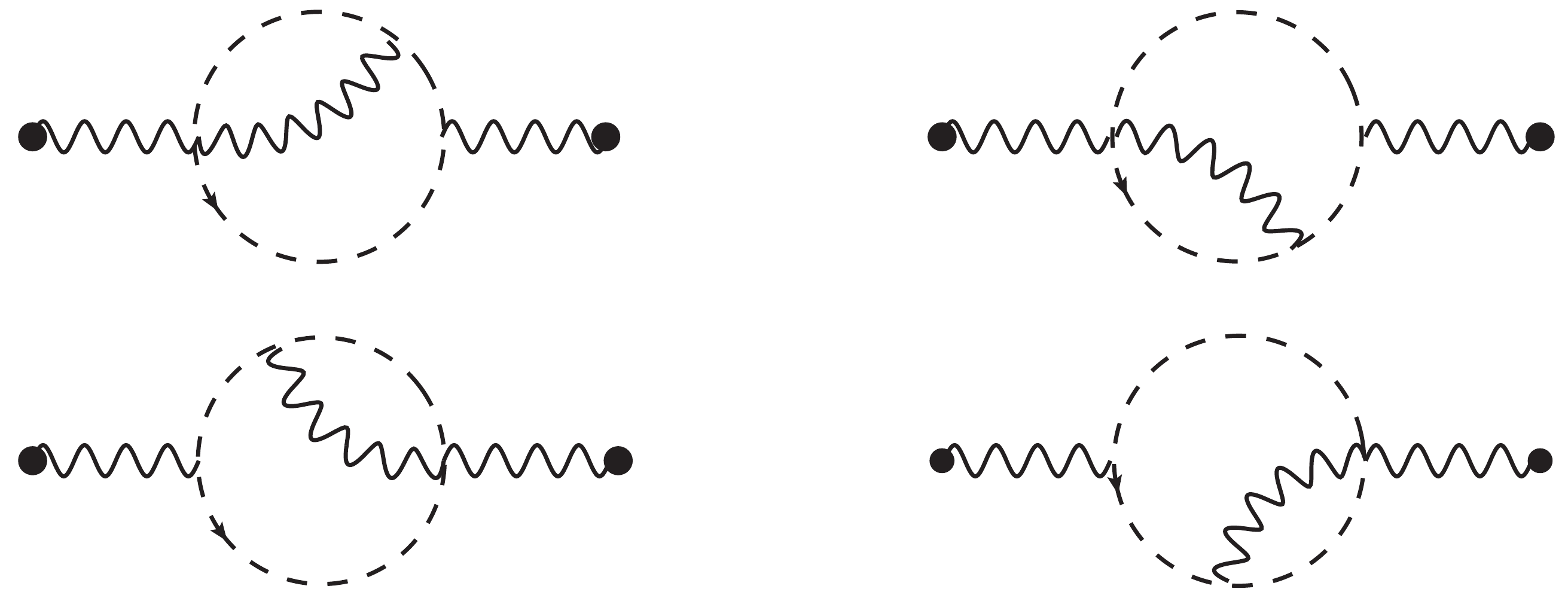}
\caption{Two loop diagram}
\label{tuu2}
\end{figure}

At two-loop level, we have eight diagrams shown in Fig. \ref{tuu2} to consider. In calculating the first diagram of Fig. \ref{tuu2}, we encounter over-lapping divergences which can be dealt with by  the Gegenbauer polynomial technique \citep{Chetyrkin:1980pr} and the integration by parts technique \citep{Chetyrkin:1981qh,Grozin:2007zz,Smirnov:2006ry}. The total contribution of these diagrams to $Z_A$, without any $\alpha$-dependence, is
\begin{equation}
Z_A^2=-\frac{e^4N_f}{128 \pi ^4 \epsilon }.
\label{2loop2}
\end{equation}
Here we want to emphasize that the $\alpha$-dependence of our calculation is cancelled completely and exactly between two-loop diagrams---not only the divergent part, but also the remainder part are cancelled at this order. This $\alpha$-independence will be used to prove the $\alpha$-indep\-e\-ndence of $Z_A$ in the ``indirect approach''.

Higher order contributions come from diagrams generated by inserting some renormalized scalar bubbles shown in Fig. \ref{xyz} into the internal photon lines of diagrams in Fig. \ref{tuu2}; all other diagrams are suppressed by a factor of $1/N_f$. Since the one-loop scalar bubble is transverse there is no $\alpha$-dependence in these higher order diagrams.
\subsection{Indirect Approach}
\label{3.2}
In the ``indirect approach'', apart from the usual vertices, we have a new vertex $AA$ to consider. Since the one-loop scalar bubble is transverse, the diagram carrying a photon chain having both insertions of this vertex and those of the scalar bubble will not contribute.

First, let's focus on diagrams with insertions only of the vertex $AA$. Note that because
\begin{equation}
D_F^{\mu\nu}(k)\tilde {D}_{\nu\rho}(k)D_F^{\rho\sigma}(k)\propto \tilde {D}^{\mu\sigma}(k),
\label{cbuj}
\end{equation}
insertions of this vertex lead to the the longitudinal form of the photon chain. In the ``direct approach'', the $\alpha$-indepen\-d\-e\-n\-ce of our calculations at two-loop level has been proven, that's is to say, the contributions from the longitudinal part of the photon propagator are cancelled exactly at two-loop level. Due to this and Eq. (\ref{cbuj}), in the ``indirect approach``, we can show that there are no effects from the vertex $AA$.

Now, we are only left with diagrams without the insertions of the new vertex $AA$ to consider. The diagrams we should consider in the ``indirect approach'', in shape, look like the corresponding diagrams in the ``direct approach'', the main difference being that the one-loop scalar bubbles and the couplings (except the two couplings attaching to the two external background legs) in the ``indirect approach'' are unrenormalized (since in this approach we don't introduce a renormalization procedure for the photon field), while usually renormalized in the ``direct approach''. Since the one-loop scalar bubble is still transverse and there is no $\alpha$-depend\-ence at the two-loop level (this follows from that in the two-loop level there is no contribution from the longitudinal part of the photon propagator), in the ``indirect approach'' there is no $\alpha$-dependence in our calculations. In what follows, we shall, in the Landau gauge, prove that a diagram (except the one-loop diagram) in the ``indirect approach'' is equivalent to a sum of an infinite number of diagrams in the ``direct approach''. To prove this, we can focus on the equivalence of the photon chain between the diagrams considered \cite{zzy2017}.

At $n+2$ loop level, the photon chain in the ``indirect approach'' takes the form
\begin{equation}
\frac{-i\big(g^{\mu\nu}k^2-k^{\mu}k^{\nu}\big)}{(-k^2)^{2+n\epsilon}}(e_0)^{2+2n}B_\epsilon^n,
\label{mgk}
\end{equation}
where we have chosen the Landau gauge and put the two bare coupling constants belonging to the two vertices linked to the photon chain in Eq. (\ref{mgk}), and $B_\epsilon$ is given by
\begin{equation}
B_\epsilon=-N_f(4\pi\mu^2)^{\epsilon}\frac{F(\epsilon)}{48\pi^2\epsilon}.
\end{equation}
To prove the equivalence, we can express the bare coupling in Eq. (\ref{mgk}) in terms of the renormalized coupling through $e_0=e/\sqrt{Z_A}$. Then, by Taylor expansion, we can rewrite Eq. (\ref{mgk}) as
\begin{equation}
\frac{\!-\!(g^{\mu\nu}k^2\!-\!k^{\mu}k^{\nu})}{(-k^2)^{2+n\epsilon}}B_\epsilon^ne^{2n+2}\Big\{1\!+\!\sum_{k=0}^{\infty}C_{n+k+1}^n(-Z_A^1)^{k+1}\Big\},
\label{moe}
\end{equation}
where we have retained only the zero order ($\mathcal {O}(1)$) terms.

In the ``direct approach'', we encounter a set of diagrams, each carrying $n$ unrenormalized scalar bubbles and a certain number of counterterms. The photon chain of a Feynman diagram of this type with $k+1$ counterterms is
\begin{equation}
\frac{-i\big(g^{\mu\nu}k^2-k^{\mu}k^{\nu}\big)}{(-k^2)^{2+n\epsilon}}e^{2n+2}B_\epsilon^n\Big(-Z_3^1\Big)^{k+1}.
\label{pnw}
\end{equation}
Since an interchange between a counterterms and an unrenormalized scalar bubble does not bring any change in the expression for the photon chain, we have equivalent diagrams in the ``direct approach''; the number of  equivalent diagrams of the type being considered is $C_{n+k+1}^n$, which is just the coefficient of $(-Z_A^1)^{k+1}$ appearing in Eq. (\ref{moe}). Multiplying Eq. (\ref{pnw}) with this combinatorial factor, taking the summation over $k$, recalling $Z_A^1=Z_3^1$, we can establish that the equivalence bet\-w\-een the ``direct approach'' and the ``indirect approach'' is proven.
\section{The beta function and the Borel transform}
\label{5}
In this section, we shall investigate the Borel transform of the two point Green's function and derive an analytic expression for the beta function by two different approaches, which here we call LTR approach and RTL approach respectively. Having shown the equivalence between the ``direct approach'' and the ``indirect approach'' and the $\alpha$-indepe\-n\-d\-ence of our calculations, in what follows we shall use the ``direct approach'' and proceed in the Landau gauge.

Before the concrete discussion, we want to say that since the similarities of Feynman rules and the identity $Z_e=1/\sqrt{Z_3}$ (or $Z_1=Z_2$) proved in \ref{APD}, the calculation of the two point Green's function in the normal field method is equivalent to that in the ``direct approach''. This also can be understood from the property of the background field method. According to the presentation of Ref. \cite{Abbott:1981ke}, the effective action we get by using the background field method and the gauge-fixing term $\mathcal {L}_{gf}(A,A_b)$ given in Eq. (\ref{gfix}) is equal to the conventional effective action calculated with the gauge-fixing term  $\mathcal {L}_{gf}(A-A_b,A_b)$ and evaluated at $A=A_b$. Since the calculation of the effective action just involves 1PI diagrams, we can neglect the terms in $\mathcal {L}_{gf}(A-A_b,A_b)$ which have only zero or one quantum photon field, that's is to say, we can calculate the conventional effective action with the usual gauge-fixing term $\mathcal {L}_{gf}(A,0)$. Therefore in what follows, our investigation about the Borel transform of the two point Green's function in the context of the ``direct approach'' of the background field method can be applied to the normal field method of scalar QED.
\subsection{A brief introduction to Borel transform}
\label{4.1}
In quantum field theory, to extend our calculation to all Feynman diagrams is difficult and beyond our calculational powers. Many successful applications of quantum field theory are based on the use of perturbation methods, and the result is usually expressed as a series:
\begin{equation}
R[g]=\sum r_ng^n.
\label{bf00}
\end{equation}
An important issue in a series is whether the series is convergent or not. For example, in some cases the coefficients $r_n$ may grow as $n!$, which indicates that this series has zero radius of convergence \cite{Weinberg:1996kr}.

There is a mathematical technique called Borel transform which can be used to improve the convergence property of a series. The Borel transform of $R[g]$, in this work, is defined by
\begin{equation}
B_R[t]=\sum \frac{r_n}{n!}t^n,
\label{bt}
\end{equation}
which obviously has better convergence properties than the original series $R[g]$. After the acquirement of $B_R[t]$, the recovering of $R[g]$ is formally done through
\begin{equation}
g R[g]=\int_0^{\infty}e^{\frac{-t}{g}}B_R[t]dt.
\label{MBR}
\end{equation}
However, if there are singularities in $B_R[t]$, we can't guarantee the validity of Eq. (\ref{MBR}).  A singularity in $B_R[t]$ is called a ultraviolet or infrared renormalon (which name you call it depends on how this renormalon appears), and the renormalon may prevent us from using Eq. (\ref{MBR}) to recover $R[g]$.
\subsection{LTR approach}
\label{4.2}
As has been said before, in the ``direct approach'', the higher order contributions come from diagrams generated by inserting some (unrenormalized) one-loop scalar bubbles and their counterterms into the internal photon lines of the two loop diagrams shown in Fig. \ref{tuu2}. The insertion of an unrenormalized scalar bubble into the photon chain of momentum $k$ gives a multiplicative factor
\begin{equation}
-g\Big(\frac{4\pi\mu^2}{-k^2}\Big)^{\epsilon}\frac{F(\epsilon)}{\epsilon},
\end{equation}
while an insertion of a counterterms does not bring any chan\-g\-e except a divergent factor $g/\epsilon$.

Now we begin our calculation with diagrams containing only $n$ unrenormalized scalar bubbles. The total expression for these diagrams is
\begin{equation}
-i(g^{{\mu}{\nu}}p^2-p^{\mu}p^{\nu})\frac{(-g)^n\pi^{\prime}[p,\epsilon,(n+2)\epsilon]}{(n+2)\epsilon^{n+1}},
\end{equation}
where $p$ is the external momentum and
\begin{equation}
\pi^{\prime}[p,\epsilon,s]=H[p,\epsilon,s]\big[F(\epsilon)\big]^{(s/\epsilon)-2}.
\label{mrc}
\end{equation}
The function $H[p,\epsilon,s]$ appearing in Eq. (\ref{mrc}) is analytic in $s$ at $s=0$ and takes the form
\begin{align}
H[p\!,\!\epsilon\!,\!u\!+\!2\epsilon]=&e^4N_f(u+2\epsilon)\Big(\frac{4\pi\mu^2}{-p^2}\Big)^{u+2\epsilon}\bigg\{\!-\!A[u,\epsilon]\Gamma
   (1\!-\!\epsilon )^2 \nonumber\\& \Gamma (1\!-\!u\!-\!2 \epsilon)\Gamma(1\!-\!u\!-\!\epsilon)\frac{\Gamma (u\!+\!\epsilon ) \Gamma (u\!+\!2 \epsilon )}{128 \pi ^4 (2 \epsilon -3)}+\nonumber\\&\!\!\frac{2 G(\epsilon,u\!-\!1)\!+\!5 G(\epsilon,u)\!+\!2 G(\epsilon,u\!+\!1)}{256 \pi ^4 (2 \epsilon -3)}\bigg\},
\label{iny}
\end{align}
where
\begin{equation}
A[u,\epsilon]=\frac{u^2 (4\epsilon\!+\!3)\!+\!u(12\epsilon ^2\!+\!\epsilon\!-\!6)\!+\!8 \big(\epsilon (\epsilon^2\!+\!\epsilon\!-\!4)\!+\!2\big)}{\Gamma (u+1)\Gamma (3-u-3 \epsilon)\Gamma (3-u-2 \epsilon)\Gamma (u+\epsilon +1)},\nonumber
\end{equation}
and the function $G(\epsilon,1+u)$ is defined proportional to
\begin{equation}
\int \frac{\ud ^d l_1 \ud ^d l_2}{(2\pi)^{2d}}\frac{1}{(l_1^2)(l_2^2)(l_3^2)(l_4^2)(l_5^2)^{1+u}},\nonumber
\end{equation}
with $l_3=l_1-p$, $l_4=l_2-p$, $l_5=l_1-l_2$ and $d=4-2\epsilon$. The appearance of $G(\epsilon,u)$ is the result of overlapping divergences we encounter in our calculation. In later subsection, we shall give more details about this function. Here the most important property of this function is that there are no poles in $G(\epsilon,1+u)$ when $u=n\epsilon$.

Eq. (\ref{iny}) can be further simplified, by means of the integration by part technique \citep{Chetyrkin:1981qh,Grozin:2007zz,Smirnov:2006ry}, to
\begin{align}
H[p,\epsilon,u\!+\!2\epsilon]&=e^4N_f(u+2\epsilon)\Big(\frac{4\pi\mu^2}{-p^2}\Big)^{u+2\epsilon}\bigg\{\frac{C[u,\epsilon]}{D[u,\epsilon]}-\nonumber\\&\frac{\big[u (u\!+\!3 \epsilon\! -\!2)+4 (\epsilon\! -\!1)\big] \text{G}\left( \epsilon ,1\!+\!u\right)}{256 \pi ^4 (2 \epsilon\!-\!3) (u\!+\!2 \epsilon\!-\!2) (u\!+\!2
   \epsilon\!-\!1)}\bigg\},
\label{iwfi}
\end{align}
where the function $C[u,\epsilon]$ is given by
\begin{align}
\label{G11}C[u,\epsilon]=&-\Gamma (1-\epsilon )^2 \Gamma (1-u-2 \epsilon) \Gamma (1-u-\epsilon) \Gamma (u+2 \epsilon )\nonumber\\&\Big\{16 (\epsilon\!-\!1)^2\big[ 2 (\epsilon
  \!-\!3) \epsilon\!+\!3\big]\!+\!u^2 \big[\epsilon  (24 \epsilon\!-\!65)\!+\!38\big]+\nonumber\\&u^3 (4 \epsilon -7)+2 u (\epsilon -1)\big[\epsilon (24 \epsilon -67)+37\big]\Big\},
\end{align}
and the function $D[u,\epsilon]$ is given by
\begin{align}
\label{G12}D[u,\epsilon]=&128 \pi ^4 (2 \epsilon -3)
   (u+2 \epsilon -2) (u+2 \epsilon -1) \Gamma (u+1)\nonumber\\& \Gamma (-u-3 \epsilon +3) \Gamma (-u-2 \epsilon +3).
\end{align}
Eqs. (\ref{iwfi}-\ref{G12}) and the analyticity of  $G(\epsilon,1+u)$ at $u=n\epsilon$ show that the function $\pi^{\prime}[p,\epsilon,s]$ ($s=n\epsilon$) is free from poles in $\epsilon$ and can be expanded in powers of $s$ and $\epsilon$:
\begin{align}
\pi^{\prime}[p,\epsilon,s]=\sum_{j=0}^{\infty}\pi_j^\prime[\epsilon]s^j,\qquad  \pi_0^{\prime}[\epsilon]=\sum_{j=0}^{\infty}c_j^{\prime}\epsilon^j.
\label{inin}
\end{align}
where, to lighten the notation, we have omitted the possible momentum-dependence of the coefficient functions $\pi_j^\prime[\epsilon]$, and $\pi_0^{\prime}[\epsilon]$ according to Eq. (\ref{mrc}) and Eqs. (\ref{iwfi}-\ref{G12}) is
\begin{equation}
\pi_0^{\prime}[\epsilon]=\frac{-e^4N_f(1-2 \epsilon ) (3-2 \epsilon ) (\epsilon -4) \Gamma (4-2 \epsilon )}{2304 \pi ^4 \Gamma (1-\epsilon ) \Gamma (2-\epsilon ) \Gamma (3-\epsilon )
   \Gamma (\epsilon +1)}.
\label{sdfin}
\end{equation}

New diagrams contributing in our approximation, can be generated by replacing some or all of the $n$ unrenormalized scalar bubbles in those diagrams by their counterterms. Taking all these diagrams into consideration, denoting the result by $\Pi_n^t(p,g)$, we have
\begin{equation}
\Pi_n^t(p,g)=\sum_{j=0}^n\frac{(-g)^n}{\epsilon^{n+1}}\frac{(-1)^j}{n+2-j}C_n^j\pi^{\prime}[p,\epsilon,(n+2-j)\epsilon].
\label{G2}
\end{equation}
where the combinatorial factor $C_n^j$ is the number of choices we own to replace just $j$ scalar bubbles with their counterterms.

Substituting  Eq. (\ref{inin}) into Eq. (\ref{G2}), we have
\begin{equation}
\Pi_n^t(p,g)=(\!-\!g)^n\bigg[\sum_{i=0}^{n+1}\frac{\pi_i^\prime[\epsilon]}{\epsilon^{n\!+\!1\!-\!i}}\sum_{j=0}^nC_n^j(\!-\!1)^j(n\!+\!2\!-\!j)^{i\!-\!1}\bigg].
\label{gbt0}
\end{equation}
Here, the sum over $i$ is truncated at $n+1$ because we are only interested in the pole terms and the finite terms. Using the following combinatorial identity\cite{PalanquesMestre:1983zy,Beneke:1992ch}
\begin{equation}
\sum_{j=0}^nC_n^j(\!-\!1)^j(n\!+\!2\!-\!j)^{i\!-\!1}=\begin{cases} 0 & (1\leqslant i\leqslant n)\\\frac{(-1)^n}{(n+1)(n+2)}& i=0\\n!&i=n+1\end{cases}\label{shyi}
\end{equation}
we can rewrite Eq. (\ref{gbt0}) as
\begin{equation}
\Pi_n^t(p,g)=\frac{g^n\pi_0^{\prime}[\epsilon]}{(n+1)(n+2)\epsilon^{n+1}}+(-g)^nn!\pi^{\prime}_{n+1}[\epsilon],
\label{gbt}
\end{equation}

Since the second term of Eq. (\ref{gbt}) suffers from no poles in $\epsilon$, the renormalization constant $Z_A$ is totally determined by $\pi_0^{\prime}[\epsilon]$. Here for later comparison with the RTL approach, we introduce a new function defined through $P(\epsilon)=\epsilon \pi^{\prime}_0[\epsilon]=\sum_nP_n\epsilon^n$, then the renormalization constant $Z_A$ is
\begin{equation}
Z_A=-\sum_{m=0}^{\infty}\sum_{n=0}^{m+1}\frac{P_ng^m}{\epsilon^{m+2-n}(m+2)(m+1)},
\label{za1}
\end{equation}
where we have used $P_0=0$. Here we want to emphasise that, through out this paper, when we talk about $Z_A$, for simplicity of our presentation, we have omitted the constant ``1'' in $Z_A$, the definition expressions in Eq. (\ref{ZA}) and Eq. (\ref{sdf}) being exceptions. The Borel transform of the two point Green's function, according to Eq. (\ref{gbt}) and its definition in Eq. (\ref{bt}), can be written as
\begin{equation}
B_{\Pi}[t]=\sum_{n=0}^{\infty}\sum_{i=0}^{n+2}\frac{{t}^n}{(n+2)!}\frac{P_i}{\epsilon^{n+2-i}}+\sum_{n=0}^{\infty}\pi^{\prime}_{n+1}[\epsilon](-t)^n,
\label{dfk4}
\end{equation}
where we have omitted the factor $-i(g^{{\mu}{\nu}}p^2-p^{\mu}p^{\nu})$, and the Borel transform of $Z_A$ can be got by extracting its pole part:
\begin{equation}
B_{Z_A}[t]=-\sum_{m=0}^{\infty}\sum_{n=0}^{m+1}\frac{P_nt^m}{\epsilon^{m+2-n}(m+2)!}.
\label{h9u}
\end{equation}

Now, we turn to the renormalized Borel transform which in the minimal subtraction scheme is obtained by subtract all the pole terms in the original Borel transform $B_{\Pi}[t]$. Thus the renormalized Borel transform, here denoted by $B_{\Pi}^0[t]$, is
\begin{equation}
B_\Pi^0[t]=\Big[\frac{B_{P}[t]}{t^2}\!-\!\frac{P_1}{t}\Big]\!-\!\Big[\frac{1}{t}\sum_{n=0}^{\infty}\pi^{\prime}_{n}[\epsilon](-t)^{n}-\frac{1}{t}\pi^{\prime}_0[\epsilon]\Big],
\label{infw}
\end{equation}
where $B_{P}[t]$ is the Borel transform of $P(x)$. When we take the limit $\epsilon\rightarrow0$, the second terms in these two brackets cancel each other, and the first term in the second bracket, according to Eq. (\ref{mrc}) and Eq. (\ref{inin}) is
\begin{equation}
-\text {exp}\Big\{t\Big[\gamma-\frac{8}{3}\Big]\Big\}\frac{H[p,0,-t]}{t}
\end{equation}
Therefore Eq. (\ref{infw}) can be combined into a compact form
\begin{equation}
B_\Pi^0[t]=\frac{B_P[t]}{t^2}-\text {exp}\Big\{t\Big[\gamma-\frac{8}{3}\Big]\Big\}\frac{H[p,0,-t]}{t}.
\label{fbr}
\end{equation}
We shall derive it again in another way and discuss it more detailedly in later subsections.
\subsection{RTL approach}
\label{RTL approach}
In previous subsection, the derivation of $Z_A$ and investigation of the Borel transform of the two point Green's functions are simplified by using the combinatorial identity shown in Eq. (\ref{shyi}). In this subsection, we will use a different approach inspired by the approach presented in Ref. \cite{Braaten:1998au} to derive $Z_A$ and discuss the Borel transform of the two point Green's function of scalar QED(before doing this work, we have used this approach in Ref. \cite{zzy2017} to study the large order behaviour of spinor QED). The essential point of this approach lies in the observation that in our approximation, to calculate the Borel transform of the two point Green's function, we can first calculate the Borel transform of the photon chain.

Obviously, the insertion of a renormalized scalar bubble into a photon chain of momentum $k$ gives a multiplicative factor
\begin{equation}
D(k^2,g)=-g\Big[\frac{F(\epsilon)}{\epsilon}\Big(\frac{4\pi\mu^2}{-k^2}\Big)^\epsilon-\frac{1}{\epsilon}\Big],
\end{equation}
from which we can write the Borel transform of this photon chain as
\begin{equation}
B_{D}^{\mu\nu}[t,k]=\text{exp}\Big\{u\Big[\frac{F(\epsilon)}{\epsilon}\Big(\frac{4\pi\mu^2}{-k^2}\Big)^{\epsilon}-\frac{1}{\epsilon}\Big]\Big\}\frac{-i(g^{\mu\nu}k^2-k^{\mu}k^{\nu})}{(k^2)^2}\
\label{bf1}
\end{equation}
where $u=-t$ and  $t$ is the Borel parameter. This can be rewritten as \cite{Braaten:1998au}
\begin{align}
B_{D}^{\mu\nu}[t,k]=K[\epsilon,u,\tilde{u}](4\pi\mu^2)^{\tilde{u}}\frac{-i(g^{\mu\nu}k^2-k^{\mu}k^{\nu})}{(-k^2)^{2+\tilde{u}}}\bigg|_{\tilde{u}=u},
\label{bf3}
\end{align}
where
\begin{equation}
K[\epsilon,u,\tilde{u}]=\text{exp}\Big\{u\Big[\frac{F(\epsilon)}{\epsilon}e^{\epsilon\frac{\partial}{\partial \tilde{u}}}-\frac{\partial}{\partial \tilde{u}}-\frac{1}{\epsilon}\Big]\Big\}.
\end{equation}
Then the Borel transform of the two point Green's function can be got by the following two steps. First we replace the photon propagators in the two loop diagrams with
\begin{equation}
\frac{-i(g^{\mu\nu}k^2-k^{\mu}k^{\nu})}{(k^2)^{2}}\Big(\frac{4\pi\mu^2}{-k^2}\Big)^{\tilde{u}},\nonumber
\end{equation}
and do the usual loop integrals. Then we operate the result of the first step with the operator $K[\epsilon,u,\tilde{u}]$, and in the end we set $\tilde {u}=u$.

The result of the first step is a very length expression of the form
\begin{equation}
-i(g^{\mu\nu}p^2-p^{\mu}p^{\nu})\frac{H[p,\epsilon,\tilde{u}+2\epsilon]}{\tilde{u}+2\epsilon},
\end{equation}
where $H[p,\epsilon,\tilde{u}+2\epsilon]$ has been given in Eq. (\ref{iny}) (or Eq. (\ref{iwfi})). The factor $-i(g^{\mu\nu}p^2-p^{\mu}p^{\nu})$ is the required polynomial in the external momentum $p$ and will be omitted later. In the second step, we can make the following trick
\begin{align}
&K[\epsilon,u,\tilde{u}]\frac{H[p,\epsilon,\tilde{u}\!+\!2\epsilon]}{\tilde{u}\!+\!2\epsilon}\bigg|_{\tilde{u}=u}=H[p,\epsilon,0]K[\epsilon,u,\tilde{u}]\frac{1}{\tilde{u}\!+\!2\epsilon}\bigg|_{\tilde{u}=u}\nonumber\\&
+K[\epsilon,u,\tilde{u}]\frac{H[p,\epsilon,\tilde{u}+2\epsilon]-H[p,\epsilon,0]}{\tilde{u}+2\epsilon}\bigg|_{\tilde{u}=u}.
\label{ixne}
\end{align}
The second term of Eq. (\ref{ixne}) is free from poles because of the analyticity of $H[p,\epsilon,\tilde{u}+2\epsilon]$ in $\tilde{u}+2\epsilon$ at $\tilde{u}+2\epsilon=0$. Therefore, we can take the limit $\epsilon\rightarrow0$ within this term and get
\begin{equation}
\text {exp}\Big\{u\Big[\frac{F(\epsilon)-1}{\epsilon}\Big]\Big\}\frac{H[p,0,u]-H[p,0,0]}{u}.
\label{lyx}
\end{equation}
As regards the denominator of the first term in Eq. (\ref{ixne}), by means of the $\alpha$ integral representation, we can write it as
\begin{equation}
\frac{1}{\tilde{u}+2\epsilon}=\int_0^{\infty}\text {exp}\{-(\tilde{u}+2\epsilon)\alpha\}\ud \alpha.
\end{equation}
Substituting this representation into the first term of Eq. (\ref{ixne}), effecting some elementary calculations, we can write the first term in Eq. (\ref{ixne}) as
\begin{align}
&H(p,\epsilon,0)\text {exp}\Big\{\frac{uF(\epsilon)-u}{\epsilon}\Big\}\frac{(uF(\epsilon)-\epsilon)}{u^2F(\epsilon)^2}\nonumber\\&+\text{exp}\Big\{-\frac{u}{\epsilon}\Big\}\frac{H(p,\epsilon,0)\epsilon}{u^2F(\epsilon)^2}.
\label{t7j}
\end{align}

The first term in Eq. (\ref{t7j}) does not suffer from poles in $\epsilon$ and therefore can be written as
\begin{equation}
\text {exp}\Big\{\frac{uF(\epsilon)-u}{\epsilon}\Big\}\frac{H[p,0,0]}{u}.
\label{byj}
\end{equation}

As regards the second term in Eq. (\ref{t7j}), the exponential factor $\text {exp}(-u/\epsilon)$ can be expanded in powers of $u/\epsilon=-t/\epsilon$, and the remainder $\epsilon$-dependent part, which is just the function $P(\epsilon)=\epsilon \pi^{\prime}_0[\epsilon]$, can be expanded in powers of $\epsilon$. These two expansions guarantee that the pole terms in the Borel transform take the required form $u^i/\epsilon^j$, as will be clear later. Effecting these two expansions, we can write the second term in Eq. (\ref{t7j}) as
\begin{equation}
\sum_{m=0}^{\infty}\sum_{n=0}^{\infty}\Big(\frac{t}{\epsilon}\Big)^m\frac{1}{m!}\frac{P_n\epsilon^n}{t^2}.
\label{ine}
\end{equation}
Here, a few remarks are in order. First, because of the absence of pole terms in expression (\ref{lyx}) and expression (\ref{byj}), the pole terms of the Borel transform only come from expression (\ref{ine}), that's is to say the renormalization constant $Z_A$ is fully determined by the function $P(\epsilon)$. Second, the function $P(\epsilon)$ is only a function of $\epsilon$ and has nothing to do with the external momentum $p$, which guarantees the momentum independence of the renormalization constant $Z_A$.

According to expression (\ref{ine}), we can write the Borel transform of $Z_A$ in the minimal subtraction scheme as
\begin{equation}
B_{Z_A}[t]=-\sum_{m=0}^{\infty}\sum_{n=0}^{m+1}\frac{t^mP_n}{\epsilon^{m+2-n}(m+2)!}.
\label{ixt}
\end{equation}
The same result has been derived before in previous subsection by the LTR approach. Having got the Borel transform of $Z_A$, we can recover $Z_A$ in various ways, such as differentiating the Borel transform $B_{Z_A}[t]$ with respect to $t$ enough times and then setting $t=0$, or multiplying $m!$ back---all these methods give the same result:
\begin{equation}
Z_A=-\sum_{m=0}^{\infty}\sum_{n=0}^{m+1}\frac{P_ng^m}{\epsilon^{m+2-n}(m+2)(m+1)},
\label{inth}
\end{equation}
which has been derived before in previous subsection by the LTR approach.

Now we turn to the renormalized Borel transform of the two point Green's function. It consists of three parts of which the first and the second parts have been given in Eq. (\ref{lyx}) and Eq. (\ref{byj}) respectively and the third part according to expression (\ref{ine}) is given by
\begin{equation}
\sum_{m=0}^{\infty}\frac{P_mt^m}{t^2m!}=\frac{B_P[t]}{t^2}.
\end{equation}
Adding all these three parts, we have
\begin{equation}
B_\Pi^0[t]=\frac{B_P[t]}{t^2}-\text {exp}\Big\{t\Big[\gamma-\frac{8}{3}\Big]\Big\}\frac{H[p,0,-t]}{t},
\label{sin}
\end{equation}
which also has been derived before in previous subsection by the LTR approach.
\subsection{The beta function}
\label{5.4}
First, let's check our result with Eq. (\ref{bgx}). According to  Eq. (\ref{zbeta}) and Eq. (\ref{oneloop}), the one-loop beta function is given by $\beta_1=eg$. Using this, we can rewrite Eq. (\ref{bgx}) as
\begin{equation}
(j+1)Z_A^{(i+1,j+1)}=g(j-1)Z_A^{(i,j)}.
\label{sdnfo}
\end{equation}
Straightforward substitution shows that our result given in Eq. (\ref{inth}) (or Eq. (\ref{za1})) satisfies this equation.

Now, we turn to the determination of $\beta (e)$. Among all the  pole terms in $Z_A$, we are only interested in the simple pole part. Extracting this from $Z_A$ given in Eq. (\ref{inth}), we have
\begin{equation}
Z_A^{(1)}=-\sum_{m=0}^{\infty}\frac{P_{m+1}g^m}{(m+2)(m+1)}.
\end{equation}
The beta function, not including the one-loop result, then can be written as
\begin{align}
\beta(e)&=e\sum_{m=0}^{\infty}\frac{P_{m+1}g^m}{m+1}\nonumber\\&=\frac{eg}{N_f}\int_0^g\frac{(3-2x)(1-2x)(4-x)\Gamma(4-2x)}{\Gamma (1-x
   )\Gamma(2-x)\Gamma (3-x ) \Gamma (1+x )}\ud x,
\label{ion}
\end{align}
where we have used the fact that the coefficients $P_n$ are proportional to $e^4$.

\begin{figure}[htb]
\centering
\includegraphics[scale=0.6]{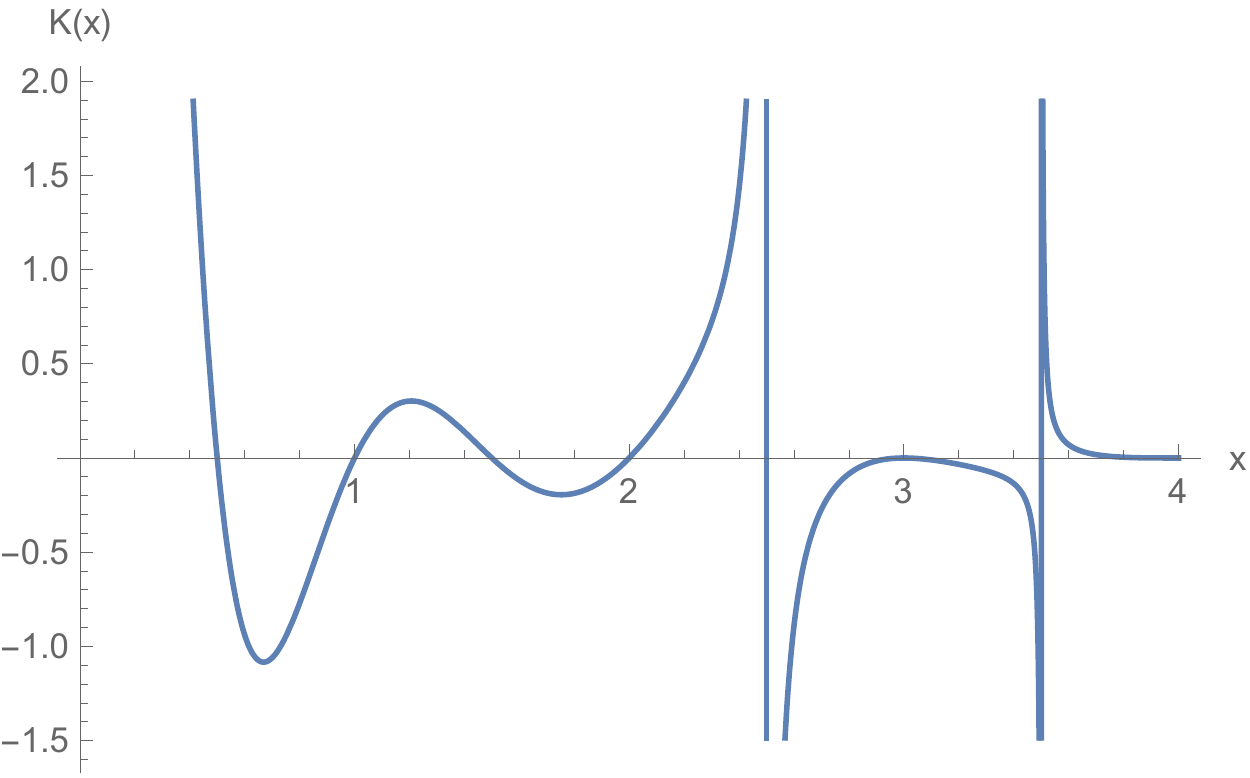}
\caption{The first two singularities of the integrand $K(x)$}
\label{555}
\end{figure}

As has been mentioned in Ref. \cite{Beneke:1998ui}, for the analyticity of a subtraction function(except at $u=0$), say $S[u]$, there is a requirement that the renormalization group functions have convergent regions, or at least they don't diverge as fast as factorials. Our result given above in Eq. (\ref{ion}) shows that up to the leading order in $1/N_f$ the beta function does have a convergent region $g<5/2$, which can be seen from the explicit expression for the integrand; here we call it $K(x)$ and depict its figure in Fig. \ref{555}. In this convergent region $g<5/2$, $\beta (e)$ is always positive. When $g$ approaches $5/2$, the beta function encounters the first logarithmic singularity.
\subsection{Renormalons}
\label{5.5}
In the renormalized Borel transform given in Eq. (\ref{sin}) (or Eq. (\ref{fbr})), the second term which here we call $D[t]$, according to Eq. (\ref{iny}), can be written as
\begin{equation}
D[t]=K(t)\Big\{S(t)\!-\!2 G(-t\!-\!1)\!-\!2 G(1\!-\!t)\!-\!5 G(-t)\Big\},
\label{ibe}
\end{equation}
where
\begin{align}
K(t)&=\text {exp}\Big\{t\Big[\gamma-\frac{8}{3}\Big]\Big\}\Big(\frac{4\pi\mu^2}{-p^2}\Big)^{-t}\frac{e^4N_f}{768 \pi ^4},\\
S(t)&=\frac{6 t^2+12 t+32}{ t^2 (t+1)^2 (t+2)^2},
\end{align}
and the function G(u) is defined through $G(u)=G(0,u)$. The function $G(\epsilon,u)$ has been expressed as a double sum in Ref. \cite{Chetyrkin:1980pr} and reduced to a one-fold series in Ref. \cite{Kotikov:1995cw} . According to Eq. (2.19) of Ref. \cite{Chetyrkin:1980pr}, $G(u)$ is given by
\begin{align}
G(u)=&2\sum_{m=0}^{\infty}\sum_{n=0}^{\infty}\frac{(-1)^m\Gamma(2-u) \Gamma (m+n+u)}{ \Gamma(u)\Gamma(m+1)\Gamma
   (m+n+2) \Gamma (-m-u+2)}\nonumber\\&\bigg\{\frac{1}{(m+n+1) (m+n+u)}+\frac{1}{(n+u)
   (m+n+u)}\nonumber\\&+\frac{1}{(m+n+1) (n-u+2)}\bigg\}.
\label{gsjo}
\end{align}
Making the replacements $m=k$ and $n=l-k$, using the identity
\begin{equation}
\frac{(-1)^k\Gamma(2-u)}{\Gamma(2-u-k)}=\frac{\Gamma(k-1+u)}{\Gamma(u-1)}\qquad (k=N),
\end{equation}
we obtain the following expression for the function $G(u)$
\begin{align}
G(u)&=\sum_{l=0}^{\infty}\sum_{k=0}^l\frac{2\Gamma(l\!+\!u)\Gamma(k\!-\!1\!+\!u)}{\Gamma(u)\Gamma(u\!-\!1)\Gamma(l\!+\!2)\Gamma(k\!+\!1)}\Big\{\frac{1}{(l+1)(l+u)}\nonumber\\
&+\frac{1}{(l-k+u)(l+u)}+\frac{1}{(l+1)(l-k+2-u)}\Big\}.
\label{SDFNI}
\end{align}
This expression for $G(u)$ is convergent for Re $u<2$ and suffers from double poles at non-positive integers from individual terms of the sum; the analytic continuation of $G(u)$ to the entire complex plane is performed under the symmetry $G(1+u)=G(1-u)$ \cite{Beneke:1992ch}.

Here we give three relevant expansions of the G function (more details about this function can be founded in Ref. \cite{Beneke:1992ch}):
\begin{align}
G(1+z)&=6\zeta(3)+\mathcal{O}(z^2),\\
G(z)&=\frac{2}{z^2}+\frac{2}{z}+\mathcal{O}(z),\\
G(z-1)&=-\frac{1}{z^2}-\frac{1}{2z}+\frac{3}{2}+\mathcal{O}(z).
\end{align}
Armed with these three expressions and the symmetry $G(1+u)=G(1-u)$, we now turn to study the singularities of the renormalized Borel transform. Obviously its component $D[t]$ suffers from a singularity at $t=0$ arising from the singularities of $G(x)$ at $x=0,-1$ and the singularity of $S(x)$ at $x=0$. Near $t=0$, $D[t]$ behaves as
\begin{equation}
D[t]=-\frac{e^4N_f}{64\pi^4t}+\mathcal {O}(1).
\end{equation}
Since we have subtracted all the pole terms, in the whole renormalized Borel transform $B_\Pi^0[t]$ there should be no singularity at $t=0$. Therefore this singularity in $D[t]$ should be cancelled. Note that in the first term of Eq. (\ref{sin}), there is also a singularity at $t=0$---near $t=0$, this term, $B_P[t]/t^2$, behaves as
\begin{equation}
\frac{P_1}{t}+\mathcal {O}(1)=\frac{e^4N_f}{64\pi^4t}+\mathcal {O}(1).
\end{equation}
Therefore in the whole renormalized Borel transform there is no singularity in $t$ at $t=0$.

Apart from the singularity at $t=0$, $D[t]$ still suffers from two kinds of singularities. The first kind of singularities come from the singularities of the function $S(t)$ which becomes singular when $t=-1,-2$. The second kind of singularities come from the singularities of the $G$-function which becomes singular when its argument becomes an integer not equivalent to $1$.

In what follows we shall omit the regular factor $K(t)$ of $D[t]$, then, near $t=-1$, $D[t]$ behaves as
\begin{equation}
\frac{3}{128 \pi ^4 (t+1)^2}+\mathcal {O}(1).
\end{equation}
This singularity at $t=-1$ is required to disappear in spinor QED because of the absence of a gauge invariant operator of dimension two (in general the singularity at $t=-n$ is accounted for by operator of dimension $2n$  \cite{Beneke:1992ch}). Near $t=-2$, $ D[t]$ behaves as
\begin{equation}
\frac{1}{64 \pi ^4 (t+2)}+\mathcal {O}(1).
\label{DT2}
\end{equation}
When $t$ is an integer not equivalent to $-1$ or $-2$, the singularities of $D[t]$ only come from the singularities of the $G$-function. Since the singularities in the $G$ function are double poles, all these singularities are double poles.

Among the singularities of $D[t]$, the singularities at $t=n$, $n=1,2,\ldots$ are called ultraviolet renormalons since they originate from high-momentum regions of integration in the loop integrals. These singularities destroy Borel summability of the series because they are on the positive real axis. The singularities at $t=n$, $n=-1,-2,\ldots$ are called infrared renormalons since they originate from low-momentum regions of integration in the loop integrals.

\section{Scheme dependence}
\label{6}
Our presentations given above is based on the adoption of the minimal subtraction scheme. In this section, we want to generalise our study to arbitrary minimal subtraction-like (MS-like) schemes such as $\overline{MS}$.

The main difference between the $\overline{MS}$ scheme and the minimal subtraction scheme is that after the subtractions of pole terms we can still subtract some finite terms in the $\overline{MS}$ scheme. And the renormalization constant $Z_A$, in the $\overline{MS}$ scheme, is usually of the form \cite{Collins:2011zzd}
\begin{equation}
Z_A=1+\sum_{n=1}^{\infty}(S_{\epsilon}g)^n\sum_{i=1}^n\frac{Z_{n,i}}{\epsilon^i},
\label{sdf}
\end{equation}
where $S_{\epsilon}$ is chosen to be of the form $S_{\epsilon}=1+a\epsilon+\mathcal {O}(\epsilon^2)$. In the $\overline{MS}$ scheme the beta function is still determined by the simple pole part of $Z_A$ and is given by
\begin{equation}
\beta(e)=-e\sum_{n=1}^{\infty}ng^nZ_{n,1}.
\label{zzbtt}
\end{equation}
In the $\overline{MS}$ scheme the finiteness of the beta function still allows us to determine the higher order poles of $Z_A$ from its single poles.

Our method described in Sect. \ref{RTL approach} can be applied to the $\overline{MS}$ sche\-m\-e with a few changes. The first change appears in the multiplicative factor $D(k^2,g)$;
\begin{equation}
D(k^2,g)\rightarrow D^\prime(k^2,gS_{\epsilon})=-g^{\prime}\Big\{\frac{F^\prime(\epsilon)}{\epsilon}\Big(\frac{4\pi\mu^2}{-k^2}\Big)^\epsilon-\frac{1}{\epsilon}\Big\},
\end{equation}
where
\begin{equation}
F^\prime(\epsilon)=F(\epsilon)/S_{\epsilon},\qquad \qquad\qquad g^{\prime}=gS_{\epsilon}.
\end{equation}
This change leads to the change: $P(\epsilon)\rightarrow P^{\prime}(\epsilon)=S_{\epsilon}^2P(\epsilon)$.

As has been shown in Sect. \ref{RTL approach}, we begin our discussion about the Borel transform of the two point Green's function with the Borel transform of the photon chain. Therefore, analogous to the situation in the minimal subtraction scheme, in the  $\overline{MS}$ sche\-m\-e, there are two coupling constants in the renormalization constant $Z_A$, which should not be regarded as coupling constants related to the Borel parameter $t$. By this we mean that if we write
\begin{equation}
Z_A(g^{\prime})=\sum_n Z_n(\epsilon)(g^\prime)^{n+2},
\label{8n1}
\end{equation}
where $Z_n(\epsilon)$ consists of terms of poles in $\epsilon$,
then the Borel transform of it is
\begin{equation}
B_{Z_A}[t]=S_{\epsilon}^2\sum_n\frac{g^2Z_n(\epsilon)t^ n}{n!}.
\label{insy}
\end{equation}
According to our procedure presented in Sect. \ref{RTL approach}, in the $\overline{MS}$ sche\-m\-e the third part of the Borel transform of the two point Green's function which suffers from poles in $\epsilon$ is
\begin{equation}
\text {exp}\Big(\frac{t}{\epsilon}\Big)\frac{H(p,\epsilon,0)\epsilon}{t^2F^{\prime}(\epsilon)^2}=S_{\epsilon}^2\text {exp}\Big(\frac{t}{\epsilon}\Big)\frac{H(p,\epsilon,0)\epsilon}{t^2F(\epsilon)^2}
\end{equation}
which, by expansion, can be written as
\begin{equation}
S_{\epsilon}^2\sum_{m=0}^{\infty}\sum_{n=0}^{\infty}\Big(\frac{t}{\epsilon}\Big)^m\frac{P_n\epsilon^n}{m!t^2}.
\label{pvr}
\end{equation}
Therefore comparing Eq. (\ref{insy}) with Eq. (\ref{pvr}), we in the $\overline{MS}$ scheme have
\begin{equation}
Z_A(g)=-\sum_{m=0}^{\infty}\sum_{n=0}^{m+1}\frac{P_nS_{\epsilon}^{m+2}g^m}{\epsilon^{m+2-n}(m+2)(m+1)},
\end{equation}
which indicates that the coefficients $Z_{n,i}$ ($n\geqslant2$) appearing in Eq. (\ref{sdf}) is scheme-independent within the $MS$-like schemes (here we should point out that the coefficients $P_n$ have a factor $e^4$ and can provide the required $g^2$). Having this and Eq. (\ref{zzbtt}) in mind, we can establish that in all the $MS$-like schemes the beta functions are the same beta function.

The finite part of expression (\ref{pvr}) is
\begin{equation}S_{\epsilon}^2\sum_{m=0}^{\infty}\frac{P_mt^m}{m!t^2}=\frac{S_{\epsilon}^2}{t^2}B_P[t]=\frac{B_{P}[t]}{t^2},
\end{equation}
where we have used $\lim_{\epsilon\to0}S_{\epsilon}=1$. Then along our procedure presented in Sect. \ref{RTL approach}, we can write the renormalized Borel transform in the $\overline{MS}$ scheme as
\begin{equation}
B_\Pi^0[t]=\frac{B_{P}[t]}{t^2}-\text {exp}\Big\{-t\Big[\frac{F^\prime(\epsilon)-1}{\epsilon}\Big]\Big\}\frac{H[p,0,-t]}{t},
\label{intc}
\end{equation}
Since the scheme dependence of $F^{\prime}(\epsilon)$, the renormalized Borel transform $B_\Pi^0[t]$ given in Eq. (\ref{intc}) is scheme-dependent. This is not surprising, since the finite results obtained by renormalization can be changed by changing the renormalization scheme and the Borel transform is defined as a Borel transform with respect to a renormalized coupling $g^{\prime}$ which is also scheme-dependent. Obviously, when $t$ approaches 0, the renormalized Borel transform doesn't suffer from a singularity. When we change our scheme by changing $S_{\epsilon}$, the only change is the change in the argument of the exponential function in the second term of Eq. (\ref{intc}), which depends only on $a$, the $\epsilon$ part of $S_{\epsilon}$. This property is in accordance with a general property of the $\overline{MS}$ scheme that we have various choices for $S_{\epsilon}$, but only the $\epsilon$ part of $S_{\epsilon}$ affect the renormalized Green's function \cite{Collins:2011zzd}. Also since the locations of the renormalons are determined by $H[p,0,-t]/t$, when we change our $\overline{MS}$ scheme we don't change the locations of the renormalons.
\section{The numerical and analytic values of the beta function}
\label{7}
The beta function $\beta(e)$ not including the one-loop result has been given in Eq. (\ref{ion}), where the coefficients $P_n$ can be got by expanding $P(x)$ in powers of $x$. Here, we give the first terms of $\beta(e)$ by doing this expansion
\begin{align}
&\beta(e)=eg+\frac{eg^2}{N_f}\bigg\{36-\frac{147 g}{2}+\frac{323 g^2}{6}+\frac{g^3}{4} \Big[72 \zeta (3)-\frac{113}{4}\Big]\nonumber\\&+\frac{g^4}{5} \Big[-294 \zeta (3)+\frac{39}{8}+\frac{6 \pi ^4}{5}\Big]+\frac{g^5}{6} \Big[323 \zeta
   (3)+216 \zeta (5)+\nonumber\\&\frac{87}{16}-\frac{49 \pi ^4}{10}\Big]+\frac{g^6}{7} \Big[-\frac{113 \zeta (3)}{2}+72 \zeta (3)^2-882 \zeta (5)+\frac{183}{32}\nonumber\\&+\frac{323
   \pi ^4}{60}+\frac{8 \pi ^6}{21}\Big]+\frac{g^7}{8}\Big[\frac{39 \zeta (3)}{4}-294 \zeta (3)^2+\frac{\pi ^4}{120}\times\nonumber\\& (288 \zeta (3)-113)+969 \zeta (5)+648
   \zeta (7)+\frac{375}{64}-\frac{14 \pi ^6}{9}\Big]+\nonumber\\&\frac{g^8}{9}\Big[\frac{13}{80}\pi ^4 -\frac{49 \zeta (3)}{5}\pi ^4+323 \zeta (3)^2-\frac{339 \zeta
   (5)}{2}+\frac{87}{8}\zeta (3)+\nonumber\\&432 \zeta (5)\zeta (3)\!-\!2646 \zeta (7)\!+\!\frac{759}{128}\!+\!\frac{323 \pi ^6}{189}\!+\!\frac{7 \pi ^8}{50}\Big]\bigg\}\ldots
\label{xpm2}
\end{align}
Our result given above up to three loop level is in agreement with the results given in the literature \cite{Dunne:2001pp,Chetyrkin:1983qc,Pickering:2001lfn}. In Table. \ref{tttt}, we give some numerical results for the coefficients $P_n$.
\begin{table*}[htb]
\begin{center}
    \begin{tabular}{cccccc}
       \hline\hline
        n & 1 & 2 & 3 & 4 & 5 \\
       \hline\hline
       $P_n$  & $36$ & $-147$ & $161.5$ & $58.2981$ & $-231.639$ \\
              \hline\hline
      n & 6 & 7 & 8&9 & 10\\
\hline\hline
       $P_n$  &$140.374$ & $17.8971$ & $-55.242$ & $20.0692$ & $2.81043$ \\
           \hline\hline
       n & 11 &12 &13&14 & 15\\
        \hline\hline
  $P_n$  &$ -4.01282 $& $0.874324$ & $0.172297$ & $-0.119144$ & $1.47442\times10^{-2}$ \\
         \hline\hline
     n & 16 & 17 & 18& 19 & 20\\
\hline\hline
       $P_n$  &$4.4666\times10^{-3}$ & $-1.7065\times10^{-3}$ & $9.7507\times10^{-5}$ & $5.70104\times10^{-5}$ & $-1.30831\times10^{-5}$ \\
           \hline\hline
     \end{tabular}
\end{center}
\caption{Numerical results for $P_n$ (with the factor$e^4 N_f/2304\pi^4$ being ommited). The first three results are exact values.}
\label{tttt}
\end{table*}

In expression (\ref{xpm2}), there are only rational numbers and the riemann zeta functions $\zeta(s)$. This property can be understood from the following relationship between the riemann zeta functions and the $\Gamma$ functions
\begin{equation}
\Gamma(1+z)=\text {exp}\Big\{-\gamma z+\sum_{k=2}^{\infty}\frac{\zeta (k)}{k}(-z)^k\Big\} \quad(\mid z\mid<1),
\label{xot}
\end{equation}
and the following relationship between $\pi^{2n}$ and $\zeta (2n)$
\begin{equation}
\zeta (2n)=\frac{(-1)^{n+1}B_{2n}(2\pi)^{2n}}{2(2n)!},
\end{equation}
where the Bernoulli numbers $B_{2n}$ are a sequence of rational numbers.

Note that all the $\Gamma$ functions in $P(\epsilon)$ are of the form $\Gamma(a+b\epsilon)$ ($a$ and $b$ being integers), and therefore can be chan\-g\-ed into the standard form $\Gamma(a+b\epsilon)=C[a,\epsilon]\Gamma(1+b\epsilon)$ where $C[a,\epsilon]$ is a polynomial in $a$ and $\epsilon$. Doing these standard transformations for all the $\Gamma$ functions in $P(\epsilon)$, using formula (\ref{xot}), we can establish that the Euler constant $\gamma$ doesn't enter into our expression for the beta function.
\section{Discussion and Conclusion}
\label{8}
In this paper, we in the large $N_f$ approximation have calculated the beta function of scalar QED at the first nontrivial order in $1/N_f$ by two different ways. We have derived an analytical expression with a finite radius of convergence for the beta function. In the convergent region $g<5/2$, the beta function is always positive which indicates that in this region there are no nontrivial fixed points arising from the zeroes of the beta function. Scheme dependence issues also have been discussed. We have shown that the beta function is scheme-independent in  $MS$-like schemes, while the renormalized Borel transform suffering from ultraviolet renormalons at $t= n$ and infrared renormalons at $t=-n$ ($n=1,2\ldots\ldots$), is scheme dependent. Furthermore, we have made clear the role played by the gauge parameter by carrying out its renormalization in both approaches (the ``direct'' approach and the ``indirect approach'') of the background field method, and the equivalence between these two approaches has been proven.

The RTL approach we used in Sec. \ref{RTL approach} can be generalized to other theories, such as Yukawa theory and Yang-Mills theory. Its generalization to Yukawa theory at the leading order in $N_f$ is straightforward. When we extend this to Yang-Mills theory some new features appear. In Yang-Mills theory, we encounter three-gauge-boson vertices, and the vertex graph with a fermion loop and three external gauge fields doesn't vanish. Therefore even at the leading order in $1/N_f$ we have to deal with diagrams with two bubble chain insertions. In this case we can use the fact that the Borel transform of a product of series is a convolution, that is to say, at this order and even higher order we can replace each bubble chain in the diagram considered with its Borel transform, do the usual loop integrals and in the end do the convolution integral (more details can be found in Ref. \cite{Beneke:1998ui}).

Finally, let's turn to the singularity structurer of Eq. (\ref{ion}) which has implications for the existence of nontrivial fixed points in the beta function and makes a contribution to better understand the asymptotic behaviour of scalar QED. Obviously the integrand $K(x)$ of expression in Eq. (\ref{ion}) suffers from poles at $x=5/2+n$ $(n=0,1\ldots)$. The appearance of these poles is due to the singularity of $\Gamma(4-2x)$ at  $x=5/2+n$ $(n=0,1\ldots)$ and therefore leads to logarithmic singularities of the beta fun\-c\-tion at $g=5/2+n$ $(n=0,1\ldots)$, which usually can be dealt with by Cauchy principal value prescription \cite{Holdom:2010qs,Alanne:2018ene}. Assuming this prescription, it can be shown that there are a UV fixed point at $g\lesssim7/2$ and a symmetric IR fixed point at $g\gtrsim7/2$ (with the explicit value depending on the value of $N_f$). The same quantitative analysis can be extended to other poles at  $g=7/2+n$ $(n=1,2\ldots)$.

\begin{acknowledgements}
One of us (Z.Y. Zheng) is deeply indebted to Prof.Y. Q. Chen for numerous discussions and helpful suggestions. The work of Z. Y. Zheng is Supported by Key Research Program of Frontier Sciences, CAS, Grant No. QYZDY-SSW-SYS006. The work of G. G. Deng is partially supported by the National Natural Science Foundation of China(11733001, U1531245, 11590782).
\end{acknowledgements}
\appendix
\section{A brief prof of $Z_{\alpha}=Z_3$ and $Z_e=1/\sqrt{Z_3}$}
\label{APD}
In this appendix, following a presentation given in Ref. \cite{book:949460} to prove the identities $Z_e=1/\sqrt{Z_3}$ and $Z_3=Z_{\alpha}$ in spinor QED, we give a standard proof of these two identities in scalar QED.

First we change the integration variables in the functional integral according to the following transformations
\begin{align}
A^{\mu}(x)&\Rightarrow A^{\mu}(x)+\frac{1}{e_0}\partial ^{\mu}\rho(x),\\
\phi (x)&\Rightarrow e^{-i\rho(x)}\phi(x).
\end{align}
The measure of the functional integral is invariant under these transformations. Thus according to the invariance of the functional integral under changing the integration variables, we have
\begin{align}
&\int\mathcal {D}A\mathcal D[\phi]\mathcal {D}\phi^*\text {exp}\Big\{i\int\ud^4x\mathcal {L}_t\Big\}\times\nonumber\\
&\Big\{\int\ud^4x\Big[\frac{-1}{\alpha_0}[\partial_{\mu}A^{\mu}(x)]\Box\frac{\rho(x)}{e_0}+J_A^{\mu}(x) \frac{\partial_{\mu} \rho(x)}{e_0}\nonumber\\
&-i\rho(x)\phi (x)J_{\phi}(x)+i\rho(x)\phi^*(x)J_{\phi*}(x)\Big]\Big\}=0.
\end{align}
Here we retain only the first order variation in the fields, and put all terms, including the gauge-fixing term and the external source ($J_A^{\mu}, J_{\phi}, J_{\phi*}$) terms, in $\mathcal {L}_t$. After integrating by parts and eliminating $\rho(x)$ in this equation, we arrive at
\begin{align}
&\frac{-1}{\alpha_0 e_0}\Box\partial_{\mu}<A^{\mu}(x)>_J-\frac{\partial_{\mu}J_A^{\mu}(x)}{e_0}-i<\phi (x)>_JJ_{\phi}(x)\nonumber\\
&+i<\phi^*(x)>_JJ_{\phi*}(x)=0,
\label{Gama}
\end{align}
where we have used  $<\mathcal {O}>_J$ to represent the vacuum expectation value of the operator $\mathcal {O}$ in the presence of the external source $J$. Expressing Eq. (\ref{Gama}) in terms of the effective action $\Gamma$, we have
\begin{align}
&\frac{-1}{\alpha_0 e_0}\Box\partial_{\mu}A^{\mu}(x)+\frac{1}{e_0}\partial_{\mu}\frac{\delta \Gamma}{\delta A_{\mu}(x)}+i\phi (x)\frac{\delta \Gamma}{\delta\phi(x)}\nonumber\\
&-i\phi^*(x)\frac{\delta \Gamma}{\delta \phi^*(x)}=0.
\label{Gama1}
\end{align}
Differentiating this equation with respect to $A_{\nu}(y)$ and then setting $A=\phi=\phi^*=0$, we have
\begin{equation}
\frac{-\Box\partial^{\nu}_x\delta^4(x-y)}{\alpha_0 e_0}=-\frac{1}{e_0}\partial_{\mu}\frac{\delta \Gamma}{\delta A_{\mu}(x)\delta A_{\nu}(y)}.
\end{equation}
In momentum space this equation reads
\begin{equation}
-\frac{p^2p_{\nu}}{\alpha_0}=p^{\mu}\Gamma_{\mu\nu}^{AA}(p,-p),
\label{Gama2}
\end{equation}
that's to say the higher-order contributions to photon self-energy graph are transverse. Therefore we can set $Z_3=Z_{\alpha}$. Another way to derive this identity is to go to the renormalized version of Eq. (\ref{Gama2})
\begin{equation}
-\frac{p^2p_{\nu}}{Z_{\alpha}\alpha}=\frac{p^{\mu}\Gamma_{r\mu\nu}^{AA}(p,-p)}{Z_3},
\end{equation}
which indicates that $Z_{\alpha}/Z_3$ is finite; i.e. we can set $Z_3=Z_{\alpha}$.

Differentiating Eq. (\ref{Gama1}) with respect to $\phi(y)$ and $\phi^*(z)$, setting $A=\phi=\phi^*=0$, we have
\begin{align}
-\frac{1}{e_0}\partial^{\mu}\frac{\delta \Gamma}{\delta A^{\mu}(x)\delta \phi(y)\delta \phi^*(z)}&=i\frac{\delta \Gamma}{\delta \phi(x)\delta \phi^*(z)}\delta^4(x-y)\nonumber\\
&-i\frac{\delta \Gamma}{\delta \phi(y)\delta \phi^*(x)}\delta^4(x-z).
\end{align}
In momentum space this equation reads
\begin{equation}
p^{\mu}\Gamma_{\mu}^{A\phi\phi^*}(p,q,l)=e_0[\Gamma^{\phi\phi*}(-l,l)-\Gamma^{\phi\phi*}(q,-q)],
\end{equation}
and the renormalized version of this equation is
\begin{equation}
\frac{p^{\mu}\Gamma_{r,\mu }^{A\phi\phi^*}(p,q,l)}{\sqrt {Z_3}Z_2}=\frac{e_0[\Gamma_r^{\phi\phi*}(-l,l)-\Gamma_r^{\phi\phi*}(q,-q)]}{Z_2},
\end{equation}
from which we conclude that $e_0\sqrt {Z_3}$ must be finite, i.e. we can set $Z_e=Z_3^{\frac{-1}{2}}$.

\end{document}